\documentclass[fleqn,usenatbib]{mnras}

\usepackage{newtxtext,newtxmath}

\usepackage[T1]{fontenc}

\DeclareRobustCommand{\VAN}[3]{#2}
\let\VANthebibliography\thebibliography
\def\thebibliography{\DeclareRobustCommand{\VAN}[3]{##3}\VANthebibliography}

\usepackage{graphicx}	
 
\usepackage{amsmath}	
\usepackage{amssymb}	
\usepackage{subcaption}
\usepackage{url}
\usepackage{color}
\def\red#1{\textcolor{red}{#1}}

\title{Precessing Binary Black Holes as Better Dark Sirens}

\author[Qianyun Yun et al.]{
Qianyun Yun$^{1,2}$,
Wen-Biao Han$^{3,1,4,5},$ 
Qian Hu$^{6}$,
Haiguang Xu$^{2}$
\\
$^{1}$Shanghai Astronomical Observatory,  Chinese Academy of Sciences,  Shanghai,   200030, China\\
$^{2}$School of Physics and Astronomy, Shanghai Jiao Tong University 800 Dongchuan RD.,Minhang District, Shanghai, 200240, China\\
$^{3}$Hangzhou Institute for Advanced Study, University of Chinese Academy of Sciences, Hangzhou 310124, China\\
$^{4}$ School of Astronomy and Space Science,  University of Chinese Academy of Sciences,  Beijing,  100049,  China\\
$^{5}$ Taiji Laboratory for Gravitational Wave Universe (Beijing/Hangzhou), University of Chinese Academy of Sciences, Beijing 100049, China\\
$^{6}$ Institute for Gravitational Research, School of Physics and Astronomy, University of Glasgow, Glasgow, G12 8QQ, UK 
}
\date{Accepted 2023 August 14. Received 2023 August 12; in original form 2023 June 13}

\pubyear{2023}

\begin{document}
\label{firstpage}
\pagerange{\pageref{firstpage}--\pageref{lastpage}}
\maketitle



\date{\today}

\begin{abstract}
Gravitational waves (GWs) from binary black hole mergers provide unique opportunities for cosmological inference such as standard sirens. However, the accurate determination of the luminosity distance of the event is limited by the correlation between the distance and the angle between the binary's orbital angular momentum and the observer's line of sight. In the letter, we investigate the effect of precession on the distance estimation of binary black hole events for the third-generation (3G) GW detectors. We find that the precession can enhance the precision of distance inference by one order of magnitude compared to the scenario where precession is absent. 
The constraint on the host galaxies can be improved due to the improved distance measurement, therefore the Hubble constant can be measured with higher precision and accuracy. These findings underscore the noteworthy impact of precession on the precision of distance estimation for 3G ground-based GW detectors, which can serve as highly accurate probes of the Universe.
\end{abstract}

\begin{keywords}
gravitational waves --  cosmological parameters  -- black hole mergers.
\end{keywords}

\section{Introduction}
In 2015, the Advanced Laser Interferometer Gravitational-wave Observatory (LIGO; ~\citep{Harry:2010zz} made a groundbreaking discovery in astronomy by detecting gravitational wave (GW) event GW150914~( \citealt{LIGOScientific:2014pky, LIGOScientific:2016dsl,  LIGOScientific:2016aoc, LIGOScientific:2018urg}); this was followed in 2017 by the second Observation Run (O2) detections~\citep{VIRGO:2014yos} from a binary neutron star system (BNS),  GW170817~(see, e.g., \citealt{LIGOScientific:2018urg, LIGOScientific:2018hze, LIGOScientific:2017vwq}), which ignited the emergence of multimessenger astronomy (MMA). The first multimessenger observations of this merger 
provided the first standard siren measurement of the Hubble constant, $H_0 = 69_{-8}^{+17} \ {\rm km s^{-1} Mpc^{-1}}$~(see, e.g., \citealt{Guidorzi:2017ogy, LIGOScientific:2018gmd,  Hotokezaka:2018dfi}).  

The discovery of the GW170817 was a significant breakthrough in the field of gravitational wave astronomy. When a gravitational wave event has an electromagnetic (EM) counterpart, the source redshift can be determined by EM observations and be used to infer cosmological parameters, which is referred as a `bright siren'~\citep{LIGOScientific:2019zcs}. However, for events without confirmed EM counterparts, such as the merger of binary black holes (BBHs), it is challenging to determine the host galaxies and their redshifts directly. `Dark siren' has been proposed ~(\citealt{Schutz:1986gp, chen2018two, LIGOScientific:2018gmd, Finke:2021aom, Leandro:2021qlc, muttoni2023dark, gair2022hitchhiker}) as a statistical framework to infer cosmological parameters without an EM counterpart. In the case of dark sirens, redshift information can be obtained through statistical analysis of the gravitational wave events in association with galaxy catalogs~(see, e.g., \citealt{holz2005using,macleod2008precision,cutler2009ultrahigh,taylor2012cosmology, Finke:2021aom, mancarella2022gravitational, branchesi2023science, LIGOScientific:2021aug}). However, compared to bright sirens, the constraints on cosmological parameters obtained from dark sirens are much less stringent due to the lack of redshift information. The large uncertainty in GW localization makes it difficult to identify the true host galaxy and determine the redshift~(see, e.g., \citealt{LIGOScientific:2018gmd, soares2019first, palmese2020statistical}). 


Third-generation gravitational~(3G) wave detectors~\citep{punturo2010einstein}, like the Einstein Telescope~(see, e.g., \citealt{punturo2010einstein,sathyaprakash2012scientific,maggiore2020science}) and Cosmic Explorer~(see, e.g., \citealt{reitze2019cosmic,hall2021gravitational,evans2021horizon}), will be a major improvement over the current ground-based detectors. They will have much better sensitivity and extend their low-frequency limit from 10 - 20~Hz to 1~Hz, allowing for the detection of $ 10^5-10^6$ GW signals from compact binary coalescences over a few years of observations~\citep{sathyaprakash2011scientific}. These detectors will provide a wealth of data for astrophysics, cosmology, and fundamental physics, and will allow us to explore the dark energy sector by studying the equation of state of dark energy and modified gravity theories~(see, e.g., \citealt{dalal2006short, cutler2009ultrahigh, zhao2011determination, cai2017estimating,  cai2018probing, belgacem2018gravitational, yang2019constraints,branchesi2023science, yang2023parameter}).

In order to fully exploit the potential of dark sirens in cosmology, it is crucial to accurately infer the luminosity distance and redshift of the source~\citep{gair2022hitchhiker}. However, distance-inclination angle degeneracy is a barrier to precise estimation of the source luminosity distance~($d_{L}$). A recent publication~\citep{borhanian2020dark} has demonstrated that  events with a large signal-to-noise ratio (SNR) could effectively break the degeneracy between $d_{L}$ and $\iota$. Moreover, other studies have shown that some physical effects, such as eccentricity and higher modes can help break the degeneracy in GW parameter estimation~(see, e.g., \citealt{yang2022eccentricity, yang2023parameter,ng2023_MeasuringPropertiesPrimordial}). \cite{Raymond:2008im} has shown that precession helps break degeneracies in source sky direction for a two LIGO-like detector network.  \citep{Raymond:2008im,green2021identifying} also pointed out that precession can improve uncertainties in distance estimates. In this letter, we present evidence that the precession of the compact binaries plays a  role in determining the distance for dark sirens for the 3G detectors. We anticipate that 3G detectors will be capable of limiting the number of host galaxies for precessing binary systems to a smaller range, whereas for non-precessing binaries, the number could potentially be 2-10 times greater.

The presence of precession sheds light on the formation of compact binaries. BBHs can form via two main mechanisms: common envelope evolution~\citep{kalogera2000spin,gerosa2018spin,steinle2022signatures} and dynamical capture~\citep{benacquista2013relativistic,rodriguez2016illuminating}. In the case of dynamical capture, the black hole spins are randomly oriented~\citep{benacquista2013relativistic,rodriguez2016illuminating}. The presence of precession sheds light on the formation of compact binaries and also offers an opportunity for testing black holes~\citep{zhang2022gravitational}. Precessing spins are predicted for BBHs formed in dense clusters, but can also occur in isolated sources due to supernova kicks. Using precessing BBHs as dark sirens requires accounting for theoretical uncertainties in their number and distribution. Their spin precession, influenced by factors like tides, winds, and high-mass binary star accretion, can be induced with >20 per cent accretion of their companion's envelope. Even 2 per cent accretion from a Wolf-Rayet star as the accretor can cause significant precession~\citep{steinle2022signatures}. Additionally, hierarchical mergers of stellar-mass black holes are expected to have spins clustered around ~0.7~\citep{Gerosa:2021mno}.

The goal of this letter is to investigate if spin precession in a BBH system can increase the precision of distance estimations for the parameter estimation of BBH signals. This letter endeavors to scrutinize BBH sources situated within a luminosity distance span of 3,000 to 40,000 Mpc (redshift $z$ = 0.5 - 5). We use $d_{\mathrm{L}}(z)=c(1+z) \int_0^z \frac{d z^{\prime}}{H\left(z^{\prime}\right)}$ to converse luminosity distance to redshift z. The $H_0$ is 67.7 km / (Mpc$\cdot$ s). The $\Omega_m$ is 0.3085. The $\Omega_{\Lambda}$ is 0. The  $\Omega_k$ is -1. As illustrated in \citep{frieman2008dark}, during epochs marked by a redshift range exceeding 0.5, the Universe is pre-dominantly governed by matter. In contrast, redshifts below approximately 0.5 signify the dominance of dark energy. As a result, observing and analyzing BBH sources at these particular redshifts  provides us with a precious understanding of the shift from a matter-dominated to a dark energy-dominated phase in the Universe.

\section{Methodology}
We employ the IMRPhenomPv3 model to produce precessing waveforms. IMRPhenomPv3 model takes the effects of two independent spins in the precessing dynamics into account~\citep{khan2019phenomenological}. This waveform approximant is available in {\sc LALSuite}~\citep{2020ascl.soft12021L}. 
To create the precessing waveform, the first step of  IMRPhenomPv3 is to express the two polarizations of the gravitational wave ($h_+$ and $h_{\times}$) as a linear combination of spherical harmonics with spin-weight -2~(see, e.g., \citealt{pan2014inspiral, hannam2014simple}). The GW multipoles $(h_{l,m}$) are the coefficients of these basis functions, calculated by Equation.~\ref{fuction1}. This is done in a frame aligned with the total angular momentum, and the direction of propagation is determined by the spherical polar coordinates $(\theta, \phi )$.

\begin{equation}
h(t ; \theta, \phi)=h_{+}-i h_{\times}=\sum_{\ell \geqslant 2, m} h_{\ell, m}(t) Y_{\ell, m}^{-2}(\theta, \varphi). \label{fuction1}
\end{equation}

The IMRPhenomPv3 expresses the waveforms of precessing BBHs $h^{\rm prec}_{l,m}$ as a function of the multipoles from non-precessing BBHs $h^{\rm non-prec}_{l,m}$. We only include the dominant $\ell=2$ modes, i.e.,~\citep{khan2019phenomenological}
\begin{equation}
h_{2, m}^{\text {prec }}(t)=e^{-i m \alpha(t)} \sum_{\left|m^{\prime}\right|=2} e^{i m^{\prime} \epsilon(t)} d_{m^{\prime}, m}^2(-\beta(t)) h_{2, m^{\prime}}^{\text {non-prec }}(t).
\end{equation}

Fisher matrix formalism is the common tool for forecasting detector capabilities, however, it could lose validity in precessing binaries due to the unconstrained spin angles and ill-behaved matrices~\citep{Borhanian:2022czq}. Therefore, we utilize the Bayesian parameter estimation package {\sc Bilby}~\citep{ashton2019bilby} to estimate the source parameters and compare the localization inference capabilities of the different spins. The standard Gaussian noise likelihood function $L$ is used to analyze the strain data $d_{k}$ based on the source parameters:
$\theta$~(see, e.g., \citealt{ashton2019bilby,romero2020bayesian}):
\begin{equation}
\ln \mathcal{L}(d \mid \theta)=-\frac{1}{2} \sum_k\left\{\frac{\left[d_k-\mu_k(\theta)\right]^2}{\sigma_k^2}+\ln \left(2 \pi \sigma_k^2\right)\right\}.
\end{equation}
In the present work, we employ IMRPhenomPv3 to generate the waveform $\mu(\theta)$, where $\theta$ consists of 15 parameters, $\theta=\{\mathcal{M}_c, \rm q, d_L, \theta_{JN}, ra, dec, \psi,  a_1, a_2, \theta_1, \theta_2, \phi_{JL}, \phi_{12}, t_c, \phi_c$\}, 
$\mathcal{M}_c$ and $q$ are the chirp mass and mass ratio of the two black holes. $d_L$ is the luminosity distance. `ra' and `dec' describe the sky position of the event and $\psi$ is the polarization angle. $\rm \theta_{JN}$ is the inclination angle at the reference frequency (which is set to 10 Hz in this work) for a processing system. $a_1, a_2$ are the dimensionless spin magnitudes of two black holes. The four angles $\rm \theta_1, \theta_2,\phi_{JL},\phi_{12} $ quantify the spin misalignment of the binary, which drives the system to precess. $ t_c$ is the merging time and $\phi_c$ is the coalescence  phase. We marginalize the parameters $ t_c$ and $\phi_c$ in this letter.

\begin{figure*}
\includegraphics[width=1\textwidth]{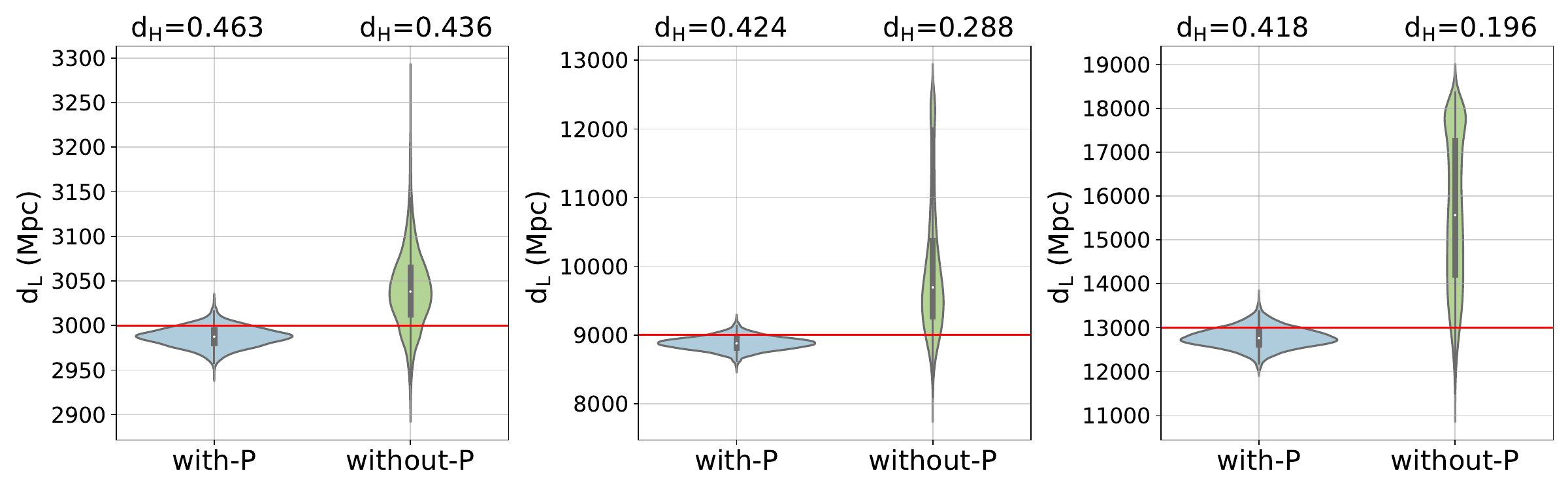} 
\includegraphics[width=1\textwidth]{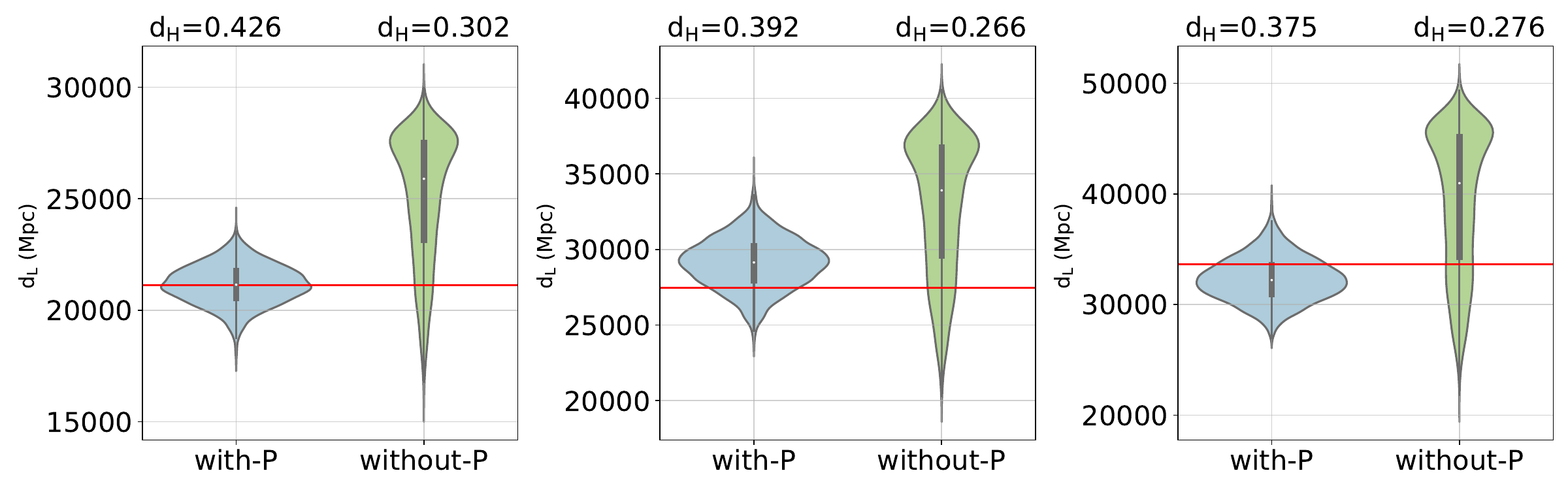} 
\caption{The posterior distributions of the luminosity distance for different injected distances.  The red line in each subfigure represents injected distance for each case. $\rm \theta_{JN}$ is $\pi/4 $ for all cases. For the precession case, $\rm \phi_{JL}$ and $\phi_{12}$ are both 0.7, respectively. The box in the middle of the violin plots represents the upper and lower quartiles of the posterior distributions.}
\label{different_l1}
\end{figure*}

We inject a signal into Gaussian noise generated with ET-D~\citep{punturo2010einstein} and CE~\citep{LIGOScientific:2016wof} design sensitivity. We assume two detectors are located at Virgo and LIGO-Hanford sites, respectively. It is important to note that this study specifically focuses on a detector network composed of ET and CE. We employ a fast sampler Nessai~\citep{williams2021nested} to generate posterior samples. Due to the inefficiency of samplers in high-SNR scenarios, where the SNR can reach values above 180 for CE detector, a significant number of computational resources will be consumed to perform complete parameter estimation.  We fix the spin parameters $\rm a_1, a_2,\theta_1, \theta_2,\phi_{JL}$ and $\phi_{12}$ to injected values in parameter estimation. The corresponding conditional posterior could be narrower than a fully marginalized posterior, but it suffices to investigate the distance-inclination degeneracy. 
The rationale for this approach is twofold: Firstly, the high SNR ratio allows us to tightly constrain the spin parameters, whereas the third Observation Run's signal-to-noise ratio effectively constrained precession~\citep{Hannam:2021pit}. Secondly, by assuming perfect knowledge of spins, we obtain conditional posterior distributions instead of marginalized posterior distributions, which may result in overconstrained error bars. Although specific numerical values might be affected, the fundamental trend of breaking the distance-inclination degeneracy due to spin effects remains valid.

We inject two types of sources with the same component masses into the detectors, one with precession and one without, to evaluate the detectors' ability to locate them and to test whether precession can alleviate the degeneracy between luminosity distance and inclination angle. The prior distribution for the luminosity distance is assumed to be uniform within the range of 1000-21000 Mpc for low redshift cases~(z = 0.5-2) and 21000-70000 Mpc for high redshift cases~(z = 2-5). We employ Hellinger distance to compute the difference between two probability distributions of priors and the estimated results. The Hellinger distance \citep{hellinger1909neue,moore2021population} between two discrete probability distributions p and q is : $\rm d_H^2=1-\sum_{i=1}^N \sqrt{p_i q_i}$. Hellinger distance is a measure of the similarity between probability distributions. 
Precessing and non-precessing systems are set at the same luminosity distance and we investigate various distances. The parameters used for injection are as follows: the chirp mass is 50; the mass ratio is 1; and $a_1, a_2$ are randomly selected from the range of 0 to 0.9.

Fig.\ref{different_l1} displays the posterior distribution of luminosity distance for the precessing and non-precessing sources at different distances.  In Fig.~\ref{different_l1}, the violin plots visually represent the distribution of luminosity  for sources at different luminosity distances. The central lines represent the box plot data, displaying the 25th, 50th, and 75th percentiles. The thin lines within the plot represent the 90 per cent confidence intervals. It is obvious that the 90 per cent confidence intervals of the distance estimation for processing BBHs are narrower compared to that of non-precessing BBHs. As depicted in Fig.~\ref{different_l1}, precession can
greatly improve the precision of luminosity distance estimation, especially at large distances. With precession, even for sources at greater distances, the luminosity distance can be estimated precisely.
In contrast, the uncertainties of $\Delta d_L $ becomes extremely large in systems without precession. Fig.~\ref{different_l1} also show the Hellinger distances for different situations.

\begin{figure}


{\includegraphics[width=0.55\textwidth]{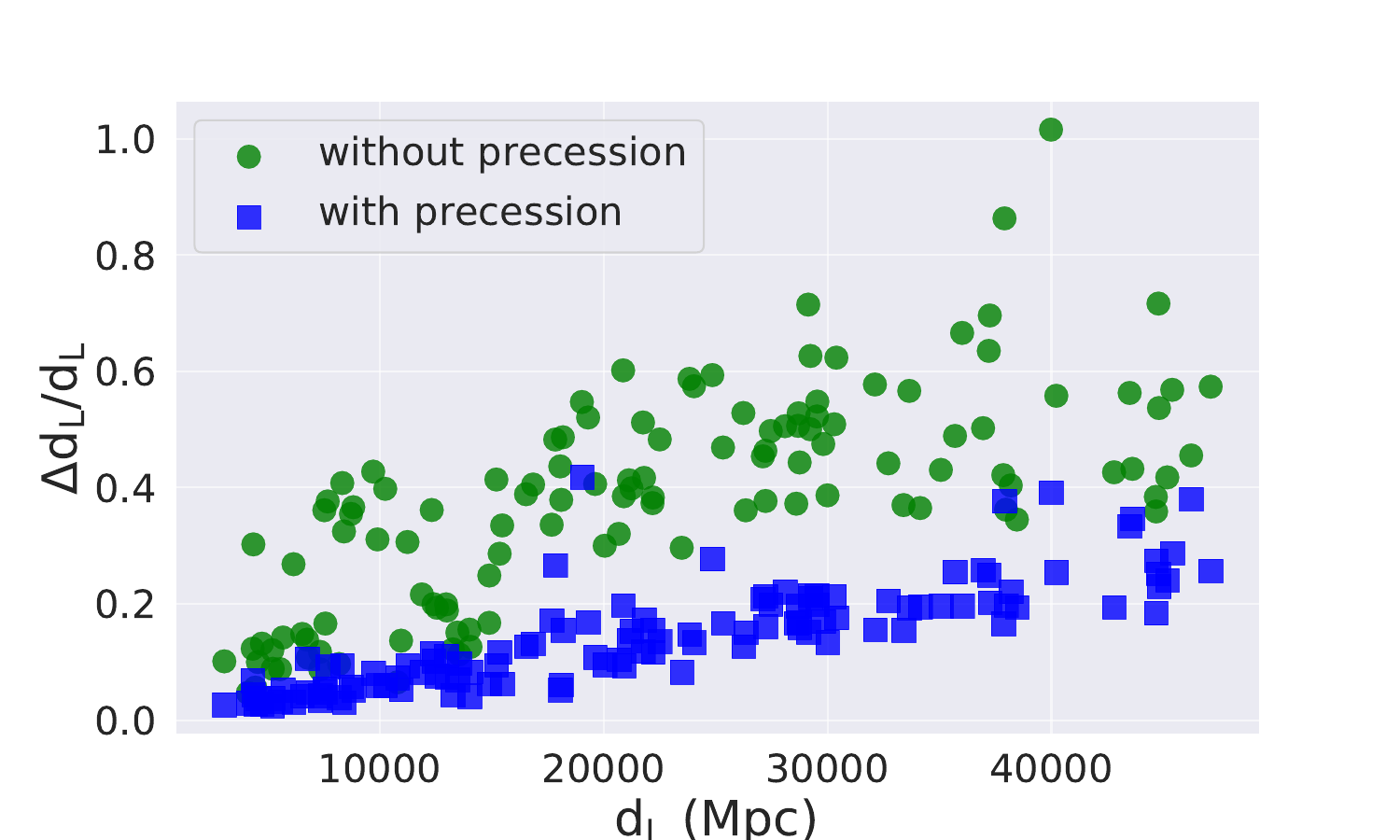}}

\caption{The confidence interval of luminosity distance divided by the luminosity distance ($\rm \Delta d_ L/ d_ L$)  for 200 random sources at different luminosity distances for both precessing cases (red dots) and non-precessing cases (blue dots).  The other parameters are set randomly.} 
\label{different_l2}
\end{figure}

In Fig.~\ref{different_l2}, we explore the differences in distance estimation between precessing and non-precessing BBHs by injecting 200 sources. These injected BBHs have randomly assigned parameters. Their luminosity distance ranges from 3000 to 48000~Mpc, $a_1$  and $a_2$ range from 0 to  0.9.
The presence of precession results in a reduction of the confidence interval for the estimated distance ($\rm \Delta d_L$). The $\rm \Delta d_L /d_L$ of precessing BBHs is consistently below 10 per cent, whereas the $\rm \Delta d_L /d_L$ of non-precessing BBHs exceeds 40 per cent. The improvement in the uncertainty of distance estimation for precessing BBHs  is more prominent at shorter distances and gradually diminishes as the redshift increases, in line with the expected decrease in SNR ratio with greater distances.


\begin{figure}
\includegraphics[width=0.5\textwidth]{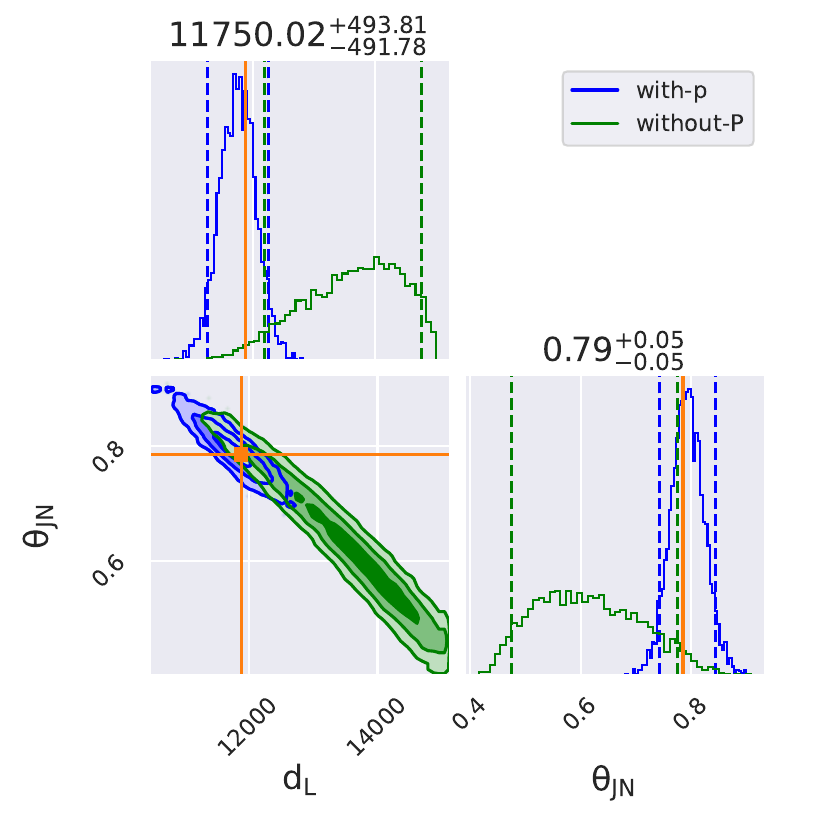} 
\includegraphics[width=0.5\textwidth]{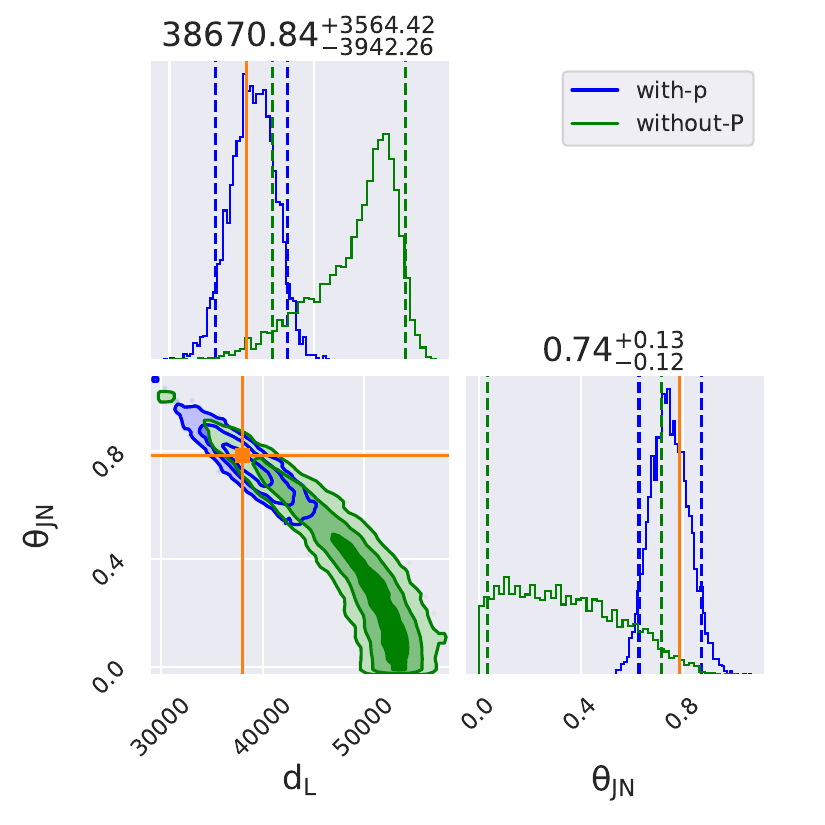} 

\caption{ The comparison of parameter estimation result of precessing BBHs and non-precessing BBHs. The injected luminosity distance is 11800~Mpc and 37974~Mpc. $\rm \theta_{JN}$ is $\pi/4 $ for both case. For the two preccesing BBHs, their $\rm \phi_{JL}$ and $\phi_{12}$ are both 1.2, respectively. Distinct and randomized sky locations are utilized for the upper and lower plots.}
\label{coner_l11800}
\end{figure}

\begin{figure}
\includegraphics[width=0.5\textwidth]{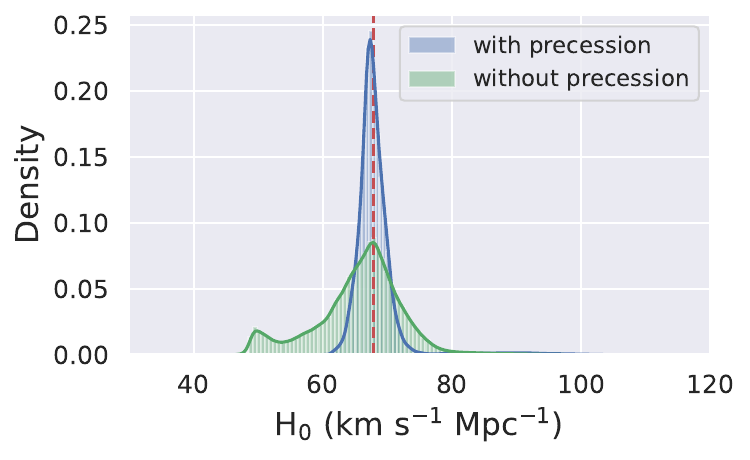}

\caption{ The comparison between the estimation of $H_0$ for dark sirens originating from precessing BBHs and those from non-precessing BBHs.}
\label{H0compare}
\end{figure}


Fig.\ref{coner_l11800} displays an example of the parameter estimation outcome for sources placed at a luminosity distance of 11800~Mpc.  
{The non-precessing system shows strong degeneracy between inclination angle and distance, which leads to an inaccurate estimation of distance.}
The precession significantly alleviates the degeneracy between $\theta_{JN}$ and the luminosity distance. The estimated distance of a precessing BBH is more precise compared to a non-precessing BBH.

Our findings highlight the  role of precession in the accurate  localization of distant gravitational wave sources. These results emphasize the importance of  precessing source  when using GW as dark sirens to measure the Hubble constant ($H_0$) in future 3G detectors.

\section{Discussion and Conclusion}
In order to utilize GW events as standard sirens for measuring the Hubble constant, it is essential to measure distances with high precision for the identification of the host galaxy. 
For the dark sirens, the strong degeneracy between the luminosity distance and the inclination angle renders it difficult to infer the distance precisely.
We emphasize the crucial role of precession in localizing gravitational wave events. By performing Bayesian parameter estimation, we demonstrate that precession significantly narrows the uncertainty of estimating luminosity distance by up to 10 times compared to non-precessing cases, especially for distant sources. This is because the precession could largely alleviate the degeneracy between distance and inclination angle.  
With the posterior localization of GW sources, one can obtain the upper and lower limits of the redshift ~(see, e.g., \citealt{chen2018two, LIGOScientific:2018gmd}), then one can filter out the candidate host galaxies. Finally, the value of $H_0$ for cosmic expansion can be obtained through the Bayesian inference~(see, e.g., \citealt{zhu2022constraining,song2022synergy}). 
The precessing BBHs can reduce the uncertainty of the luminosity distance greatly compared to the non-precessing cases, then the number of the host galaxy candidates will be reduced too as it is roughly proportional to the uncertainty of distance~\citep{fan2014bayesian}. Therefore, the measured Hubble constant will have smaller uncertainty. We make the assumption of galaxies being uniformly distributed within the co-moving volume~\citep{barausse2012evolution,wang2022hubble}. This assumption enables us to estimate the number of galaxies within the given volume. By utilizing the inferred sky positions and distance inferences, we calculate the comoving volumes in which the potential host galaxies may reside. The number density of galaxies in the Universe is 0.02 $\rm Mpc^{-3}$~\citep{wang2022hubble}.
Multiplying this density number by the volume allows us to derive the count of potential host galaxies. This calculation provides an estimation of the number of galaxies that have the potential to serve as hosts. The inclusion of precession in BBH systems can lead to a substantial decrease in the potential number of host galaxies, in comparison to non-precessing BBH systems, typically by a factor of 2-10. This reduction becomes particularly prominent when the sources are located at distances greater than 10,000 Mpc.
Additionally, we performed a simple evaluation to investigate how the precession-distance effect could influence the estimation of $H_0$   for the binary black hole sources depicted in Fig.~\ref{different_l2}. Our analysis uses the redshift of the host galaxy by converting the injected distance into a corresponding redshift value. Subsequently, we incorporated the 90~per cent interval of the estimated luminosity distance and applied the standard cosmological model to calculate the range of $H_0$. The expression for $H(z)$ was given by $H(z)=H_0 \sqrt{\Omega_{\mathrm{m}}(1+z)^3+1-\Omega_{\mathrm{m}}}$, where the equation of state (EoS) parameter of dark energy, $w$, was assumed to be -1. The values of $\Omega_m$, $\Omega_k$, and $\Omega_{\Lambda}$ remain consistent with those mentioned earlier.

The results of our calculations are illustrated in Fig.~\ref{H0compare}. The blue curve represents $H_0$ inferred from the precessing sources as dark sirens, while the green curve represents $H_0$ inferred from the non-precessing sources. It is evident from the figure that the precessing BBH sources can provide a more tightly constrained estimation of $H_0$ as dark sirens. The distributions in Fig.~\ref{H0compare} are normalized to have equal area. The areas are equal to 1.

The results in this letter suggest that precessing GW sources as dark sirens have immense potential for understanding the cosmic expansion history. By studying the expansion history, we can gain insights into the behavior of dark energy and modified gravity theories~\citep{belgacem2018gravitational}. Although our research primarily focuses on the third generation of ground-based gravitational wave detectors, the results can be extended to space-based detectors as well~\citep{lang2006measuring,lang2011measuring,klein2009parameter}. The use of precessing dark sirens could pave the way for more accurate and precise measurements of the Hubble constant and a deeper understanding of the universe's evolution.

\section*{Acknowledgements}
This work was supported by the National Key R\&D Program of China (Grant Nos. 2021YFC2203002), and the National Natural Science Foundation of China (Grant No. 12173071). Wen-Biao Han was supported by the CAS Project for Young Scientists in Basic Research (Grant No. YSBR-006).

\section*{DATA AVAILABILITY} The data underlying this article will be shared on reasonable request to the corresponding author Wen-Biao Han.

\end{document}